\documentclass{article}
\usepackage{amsmath}
\usepackage{amssymb}
\usepackage{amsthm}
\usepackage{latexsym}
\usepackage{epsfig}
\usepackage{url}

\begin{document}
\newcommand{\ontop}[2]{\genfrac{}{}{0pt}{}{#1}{#2}}

\title{Towards the Human Genotope}

\author{Peter Huggins, Lior Pachter and Bernd Sturmfels\\ Department of
Mathematics\\ University of California at Berkeley}

\date{\today}
\maketitle

 \begin{abstract}
The human genotope is the convex hull of all allele
frequency vectors that can be obtained
 from the genotypes present in the human population.
In this paper we take a few initial steps towards a description of this object,
which may be fundamental for future
population based genetics studies. 
Here we use data from the HapMap Project, 
restricted to two ENCODE regions,
to study a subpolytope of the human genotope. We study
three different approaches for obtaining informative low-dimensional 
projections of this subpolytope. The projections are specified by
projection onto few tag SNPs, principal component
analysis, and archetypal analysis. We describe the application of our 
geometric approach to  identifying structure in populations based on single
nucleotide polymorphisms.

 \end{abstract}

 \section{Introduction}
\label{Sect:Introduction}

The HapMap Project \cite{HapMap2005} is an international effort to identify the
genetic 
variation in the human population. This includes
the identification of all single nucleotide polymorphisms (SNPs) arising
in human populations. In the first major published study
 \cite{HapMap2005}, approximately ten million 
SNPs are described throughout the human genome, derived from the 
genotypes of 269 individuals from four populations. A crucial component
of the project has involved the comprehensive detection of all SNPs in ten $500$kb
regions of the human genome. The ten regions were selected from targets studied by
the ENCODE
project \cite{ENCODE2004}, whose goal is to
annotate 1\% of the human genome. Thus, while the haplotype map of the human
genome is not complete, substantial progress has been made to date, and
there is every reason to expect a completed map in the near future.

The characterization of genetic variation in the human population is
only a first step towards the fundamental goal of relating phenotypes to
genotypes. While in principle every SNP may contribute individually to
a phenotype, interactions among
loci are common. Furthermore, the problem of identifying relevant SNPs, and
understanding the
interactions among them, is confounded by the vast number of SNPs in the
genome. These issues make it non-trivial to perform {\em association
  mapping}, in which phenotypes are mapped by
analyzing genotypes of cases and controls.
Fortunately, this problem is ameliorated by two factors. First, the
number of individuals in the human population is far smaller than the
number of possible haplotypes, so that even if every individual on earth
were genotyped, the description of the
data would not involve all possible haplotypes. Secondly, there is a lot
of {\em linkage
  disequilibrium} (LD), which describes the situation in which some
combinations of alleles occur more or less frequently in a population
than would be expected by the overall frequency of the alleles in the population.
Thus, ``informative'' SNPs, also known as tag SNPs, can be identified
and used to simplify the measurement of variation, and also to reduce
the number of loci that need to be considered for association mapping. 

The data produced by the HapMap project are useful to
understand these issues. By way of example, we consider
ENCODE region ENr131. This is a 
$500064$-base region from chromosome 2. The HapMap project
has revealed that there are $2995$ SNPs in the region, meaning
that even though there are $500064$ bases, the genomes of any two
individuals differ in at most $2995$ sites. These sites typically 
contain one of 
two possible alleles, so a human haplotype is described by a binary
vector of length $2995$, and a genotype by a vector of length $2995$
whose entries are either $0,1$ or $2$. A $0$ or $2$ indicates that 
the two haplotypes agree (homozygous), and specifies the allele, and a $1$ indicates
disagreement (heterozygous). For our analysis, it is essential that $0$
and $2$ are the homozygotes. This encoding differs from the standard
encoding where $0$ and $1$ are homozygotes, and $2$ is the heterozygote
(see, e.g., \cite{Kimmel2005}).

In general, a {\em genotype} is a vector of length $n$ whose elements are from 
the alphabet $\{0,1,2\}$. The genotypes for a population of 
$k$ individuals form a matrix of size $k \times n$. In the case of ENr131, 
the HapMap project genotyped $1910$ of the $2995$ SNPs in 269
individuals, resulting in a $269 \times 1910$ matrix. In fact, it is
possible to reduce the number of SNPs that need to be considered in an
association mapping study by a factor of 10 by selecting tag SNPs.

The notion of a {\em genotope} was introduced in \cite{Beerenwinkel2006}
and further developed in \cite{Lenski2006, Debbie2006}. 
A genotope is the convex hull of all 
possible genotypes in a population. The regular
triangulations of the genotope
 describe the possible epistatic interactions among the loci. 
These objects are fundamental for analyzing linkage
disequilibrium. For example, the sign of the standard measure of LD for a pair of loci
\cite{Christiansen2000} corresponds to one of the two triangulations of
the respective genotope (the square). For the data to be examined in
this paper, the genotope is the convex hull of
$k$ points in $\{0,1,2\}^n$, so it is a subpolytope
of the $n$-dimensional cube with side length two.
In Section 2 we review the relevant mathematical theory
and we discuss the meaning of the genotope for
population genetics. 

The {\em human genotope} is the convex hull
of about $\,k  = 6.5 \cdot 10^9 \,$ points, one for each
individual in the human population, and the ambient
dimension $n$ is bounded above by the number of all SNPs.  
What can currently be derived from the HapMap data is 
a subpolytope of the human genotope which is the convex hull of only
$k = 269$ points, one for each sequenced
individual. In this paper we also restrict our attention to two
specific ENCODE regions, so that the number $n$ of informative SNPs is 
on the order of hundreds. We refer to the resulting
subpolytopes  as the {\em ENCODE genotopes}.

In Sections 4-6 we study several different low-dimensional projections of the ENCODE
genotopes.
These projections are chosen in a statistically meaningful manner, and 
we argue that the low-dimensional polytopes
are a useful geometric representation of the data. 
In Section 4 we apply Principal Component Analysis (PCA)
to determine the most significant projections of our data.
We compute the image of the ENCODE genotopes
under projection into the six most significant PCA directions. These
projections reveal the population structure of the HapMap genotypes in a
manner consistent with \cite{Price2006}.
We then contrast PCA projections with low-dimensional projections based
on tag SNPs. The resulting polytope
data are presented in Section 5. 
This geometric analysis suggests a new statistical test,
the {\em volume criterion}, for identifying informative SNPs.

In Section 6 we apply a statistical method
which is less well-known than PCA 
but possibly more informative in the population genetics context.
{\em Archetypal analysis} was
introduced by Cutler and Breiman \cite{Cutler1994} 
for identifying a small collection
of $\ell$ archetypes from $k$ given data points.
We apply this  method to our $k=269$ genotypes.
The archetypes are either genotypes
or mixtures of genotypes, so their
convex hull is a polytope with $\ell$ vertices inside an
ENCODE genotope. We call it the
{\em $\ell$-th archetope}. Its defining property is that
the total least squares error is minimal
when each genotype is replaced by its nearest 
mixture of archetypes. 
We compute various archetopes 
and explain how these may be useful for designing genetic studies.

Our studies are
preliminary and merely foreshadow the possibilities for a geometric
organization of the large amount of genotype data that are currently
being produced. We believe that low-dimensional 
projections of genotopes will be useful for correctly quantifying
population structure variation, and also for studying interaction. 
Our results
demonstrate the feasibility of computing low-dimensional projections of 
genotopes, and our analysis of the HapMap data provides a first step
towards the construction of the human genotope.

\section{The Human Genotope}

The geometric concept of a genotope was introduced in \cite{Beerenwinkel2006}
for studying epistasis and shapes of fitness landscapes. This 
model was applied in \cite{Lenski2006} to fitness data in {\em E. coli},
and its relevance for human genetics was demonstrated  in  \cite{Debbie2006}.

We briefly review the mathematical setup in \cite{Beerenwinkel2006}
but with emphasis on diploids rather than haploids.
We consider $n$ genetic loci each of which is a diploid locus
with alleles $0,1,2$. A $0$ or $2$ indicates that 
the two haplotypes agree, and specifies the allele, 
and a $1$ indicates disagreement.  The connection to
the classical genetics notation, used to discuss diploids in
\cite[Example 2.5]{Beerenwinkel2006}, is as follows:
$$ 0 \,=\, aa \,  , \quad   1 \,=\, aA = Aa \, , \quad 2 \,=\, AA. $$
There are $3^n$ genotypes, one for each element of the
set $\{0,1,2\}^n$. A population is a list of $k$ such genotypes.
It determines an empirical probability distribution on 
 $\{0,1,2\}^n$. The set of all probability distributions
on $\{0,1,2\}^n$ is a simplex of dimension $3^n-1$.
It is denoted by $\Delta$ and called the {\em population simplex}.

The vector $v \in \{0,1,2\}^n$ which represents a given
genotype records the allele at each of the $n$ sites.
If $p \in \Delta$ is a population then $p_v$ is
a number between $0$ and $1$ which indicates
the fraction of the population which has genotype $v$.
The {\em allele frequency vector} of the population $p$
is the vector $\, \sum_{v \in \{0,1,2\}^n} p_v \cdot v\,$
which lies in $\mathbb{R}^n$. The $i$-th coordinate of this vector
indicates the average number of occurrences of the lower case
letter $a$ at the $i$-th site in the population.

A {\em genotype space} is any subset $\mathcal{G}$ of $\{0,1,2\}^n$.
The elements of $\mathcal{G}$ are the genotypes that 
actually occur in some population. The genotype space $\mathcal{G}$  will always be a very small
subset of $\{0,1,2\}^n$ because the cardinality of $\mathcal{G}$
is bounded above by the number $k$ of individuals, which is
usually much smaller than $3^n$. For example, the
size of the human population, $k = 6.5 \cdot 10^9$,
is less than $3^n$ as soon as the number $n$ of sites
exceeds twenty.

We define the {\em genotope} to be the convex hull,
denoted ${\rm conv}(\mathcal{G})$, of the given
genotype space $\mathcal{G}$. Equivalently,
${\rm conv}(\mathcal{G})$ is the polytope in $\mathbb{R}^n$ which consists of
the allele frequency vectors
of all possible populations with individuals in $\mathcal{G}$.

\begin{figure}
\begin{center}
 \includegraphics[scale=0.45]{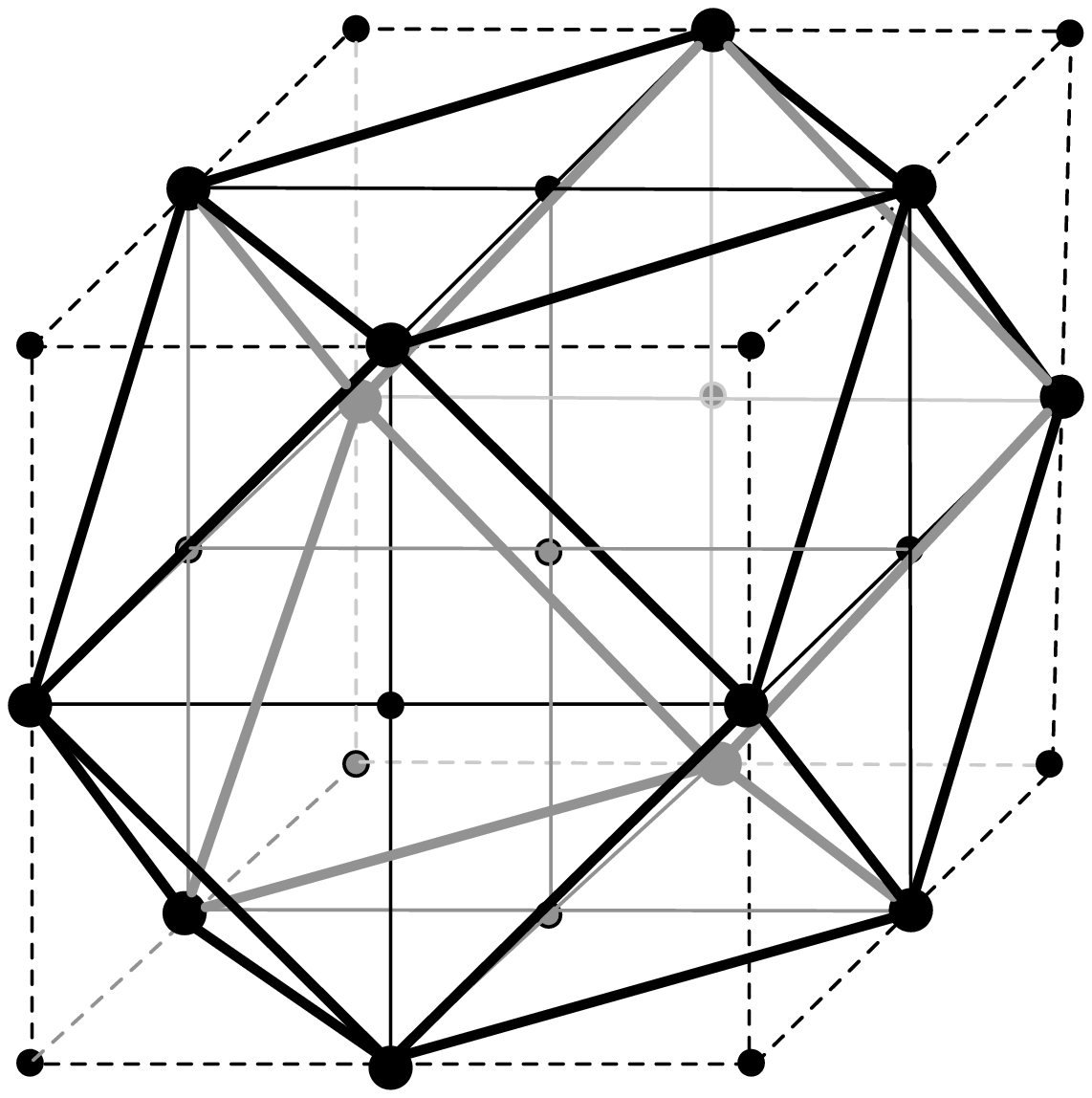}
  \includegraphics[scale=0.45]{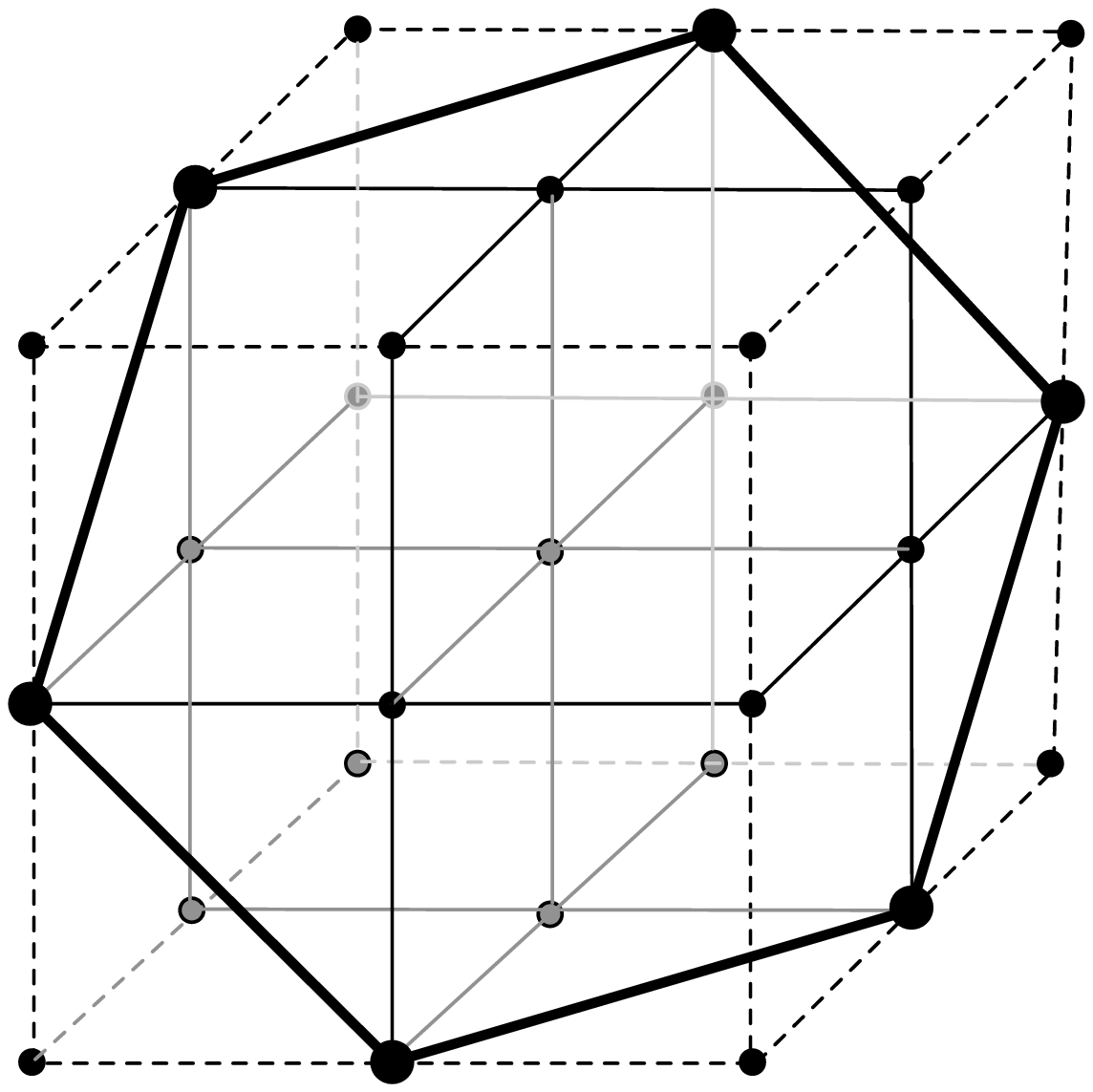}
\caption{Genotopes for three loci: a cubeoctahedron and a hexagon.}
\end{center}
\end{figure}

We illustrate the concept of a genotope for the case of
$n=3$ loci. Here $\mathcal{G}$ is any subset of
the $27$ genotypes in $\{0,1,2\}^3$, and the
genotope ${\rm conv}(G) $ is a  convex polytope
of dimension at most three.
 A basic invariant of such a polytope
is the triple $(v,e,f)$ where $v$ is the number of
vertices, $e$ is the number of edges and $f$ is the 
number of facets.
Figure 1 shows three concrete examples:
\begin{itemize}
\item If ${\cal G} = \{0,1,2\}^3$ then
the genotope $\,{\rm conv}(G)$ is the three-dimensional cube
with side lengths two. It has $(v,e,f) = (8,12,6)$.
The eight vertices correspond to the genotypes
that are homozygous at all three sites. 
\item If the eight purely homozygous genotypes
cannot occur in a population then ${\cal G}$ consists
of the remaining $19$ genotypes. Now the genotope
${\rm conv}(G)$ is a {\em cubeoctahedron}, with
$(v,e,f) = (12, 24, 14)$. Its twelve vertices correspond
to genotypes with precisely two homozygous sites.
\item If 
the alleles at all three sites must be distinct then
$$ {\cal G} \,\,\,= \,\,\, \bigl\{ (012), (021), (102), (120), (201), (210 )\bigr\}, $$
and the dimension of the genotope drops to two.
It is a regular hexagon with
$(v,e,f) = (6,6,0)$.
\end{itemize}

The theory developed in \cite{Beerenwinkel2006}  concerns
 gene interactions and 
 the shapes of fitness landscapes. By definition, a {\em fitness
landscape} is any function $\,w : \mathcal{G} \rightarrow \mathbb{R}$.
In population genetics, $w(g)$ measures the expected number of
offspring of an individual with genotype $g \in \mathcal{G}$. 
The regular triangulations of the genotope describe epistatic
interactions among the genotypes. 
In the context of human genetics, it makes sense to replace the
notion of fitness by penetrance values for a disease \cite{Ott1999} or by expression
levels of a gene \cite{Chesler2005}. A classification of 
two-dimensional genotopes
and their triangulations from this perspective is presented in \cite{Debbie2006}.

We cannot construct the human genotope at this time because the HapMap
project is not complete. However, in Section 3 we explain how the
existing preliminary HapMap data can already be used 
to reveal something about the human genotope. By restricting our 
attention to the ENCODE regions, and by taking advantage of linkage 
disequilibrium, we are able to compute
biologically meaningful low-dimensional
projections of the human genotope.

\section{Human Variation Data}

In this section we explain the data we used, how they were obtained, 
and how they were prepared.  Our data consists of 269 genotypes over
dbSNP loci in the two ENCODE regions ENr131 and ENm014 sampled by the
HapMap project.  These are the two regions listed in 
\cite[Table 8]{HapMap2005}. In 
our study, we used the non-redundant version of the
dataset which is available for download at

\begin{center}
  {\tt www.hapmap.org/genotypes/latest\_ncbi\_build34/ENCODE/}
\end{center}

\noindent
The data are grouped into four populations:
Utah-European (CEU), Han-Chinese (CHB), 
Japanese (JPT) and Yoruba (YRI).
The region ENm014 has 2315 SNPs, of which 1968 were sampled in all
four populations.
Only one of the SNPs was triallelic in the HapMap data,
but many SNPs had incomplete data due to sequencing errors.
In fact the number of SNPs
in ENm014 successfully sequenced in all 269 individuals is 790,
about a third of the total number of SNPs.
It seems that a disproportionate number of loci in region
ENm014
had missing data, evidently due to unusual 
sequencing problems in several individuals.

For simplicity we restricted our attention to the portion of
SNP loci that had two or fewer observed alleles in the HapMap data,
and which were successfully sampled in
all 269 HapMap individuals.  This implies that our projections
of the human genotope are all representable as subpolytopes of a 
standard hypercube.
After so restricting the set of loci, our data comprise 269 diploid
genotypes, over 1154 loci from region ENr131 and 790 loci from region
ENm014.  

For each SNP locus, the lexicographically smaller nucleotide was
used as the reference allele, and each of the observed 269 diploid
genotypes was encoded as a numerical genotype in $\{0,1,2\}$.
For example, for each SNP locus with nucleotide $A$ or $G$ we have $AA = 0$, $GG =
2$, and
$AG = GA = 1$.  We thereby encoded the 269 genotypes as a
$269 \times 1154$ matrix over $\{0,1,2\}$ for region ENr131 and a
$269 \times 790$ matrix for region ENm014.  The convex hull
of the rows of one of our matrices gives a projection of a subpolytope
of the human genotope.

As an illustration of our data, below is the $10 \times 15$ upper left
submatrix of the matrix for region ENr131. It encodes the first 15
out of the 1154 biallelic SNPs, which were sampled in our first 10
HapMap individuals in the CEU population:

\[
{\tt
 \left({
  \begin{tabular}{ccccccccccccccc}
  2&0&2&0&2&0&2&0&0&0&2&0&2&2&1 \\
  2&0&2&0&2&0&2&0&1&0&2&1&1&1&0 \\
  2&1&2&0&1&1&2&0&1&0&1&1&1&1&1 \\
  2&0&2&0&2&0&2&1&0&0&2&2&0&0&0 \\
  1&1&2&0&1&0&2&1&0&0&1&2&0&0&0 \\
  1&0&2&0&2&0&2&0&0&0&2&1&1&1&0 \\
  1&0&2&0&2&0&2&0&0&0&2&1&1&1&0 \\
  1&0&2&0&2&0&2&1&0&0&2&2&0&0&0 \\
  2&0&2&0&2&0&2&0&0&0&2&1&1&1&1 \\
  2&0&2&0&2&0&2&0&0&0&2&1&1&1&1 \\
  \end{tabular}
 }\right)
}
\]

Already in the above submatrix we see likely
linkage between the 5th and 6th columns,
and also between the 12th, 13th, and 14th columns.
Such covariance among SNPs will allow us to work with various
projections of genotopes, which we discuss in the following sections.
We also note that the individuals in rows 6 and 7 as
well as those in rows 9 and 10 have identical sequences in this matrix.
The eight distinct rows in this example are affinely independent,
so this projection of the human genotope into $\mathbb{R}^{15}$
is a $7$-dimensional simplex.

All of our data, along with the Linux utility used to convert the HapMap
data files into matrices can be downloaded at our supplementary website
\begin{center}
  {\tt bio.math.berkeley.edu/humangenotope/}
\end{center}
The utility takes the four population files for a particular ENCODE
region, downloaded from the above {\tt hapmap} site,
and it converts each file into the corresponding 
matrix over $\{0,1,2\}$, in both MAPLE and MATLAB format.

\section{Projections Onto Principal Components}

Principal Component Analysis (PCA) is a standard
statistical technique  for reducing the dimension of
high-dimensional data. In our study, the data is  
a $k \times n$-matrix $X $ with entries in $\{0,1,2\}$. The
 genotope under consideration is the convex hull of the
row vectors of  the matrix $X$. Our mathematical problem consists of 
applying PCA to our data in a manner that is consistent with the affine 
geometry of convex polyhedra.
For this reason, we augment the matrix $X$ 
to a $k \times (n+1)$-matrix $X'$ by adding a column
of $1$'s. 
In our situation, we have $n \geq k$, and PCA amounts to computing
the singular value decomposition
$$ X' \quad  = \quad U \cdot \Sigma \cdot V, $$
where $\Sigma$ is a  real diagonal $k \times (n+1)$-matrix whose main 
diagonal entries satisfy
$\sigma_1 \geq \sigma_2 \geq \cdots \geq \sigma_k \geq 0$.
The columns of the $k \times k$-matrix $U$
are the left singular vectors of $X'$; they form
an optimal orthonormal basis for the column space of $X'$.
Likewise, the rows of the $(n+1) \times (n+1)$-matrix $V$ are
the right singular vectors of $X'$; they are
 an optimal orthonormal basis of the row space of $X'$.
Here optimality means that 
the matrix obtained from $U \cdot \Sigma \cdot V$ by
setting  $\,\sigma_{\ell+1}  = \cdots = \sigma_k = 0\,$
is closest (in the Euclidean norm) to $X'$ among all $k \times (n+1)$-matrices of rank
$\leq \ell$.

Consider any positive integer $\, \ell \leq k$. 
Let  $\,(\Sigma \cdot V)_\ell\,$ denote the
submatrix of $\,\Sigma \cdot V\,$ consisting
of the first $\ell$ columns. We define the {\em $\ell$-th
PCA projection} of the genotope represented by $X$
to be the convex hull in $\mathbb{R}^\ell$ of the
row vectors of the matrix  $(\Sigma \cdot V)_\ell$.
This polytope can be regarded as the statistically
most significant orthogonal projection into
$\mathbb{R}^\ell$  of the given genotope  in $\mathbb{R}^n$.

\begin{figure}
\begin{center}
  \includegraphics[scale=0.65]{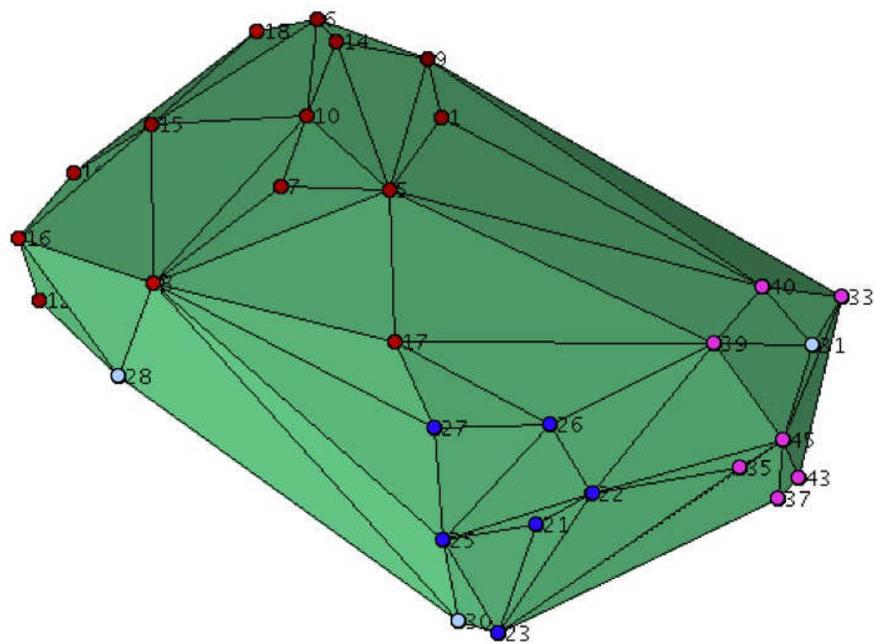}
\caption{The 3-dimensional PCA projection of the ENr131 ENCODE
  genotope. The colors of the vertices represent the different populations: red=CEU,
blue=CHB, cyan=JPT, pink=YRI.}
\end{center}
\end{figure}

The numerical computation of PCA projections is straightforward, 
equivalent to computing the SVD of the data matrix.
Using MATLAB on a Pentium 4 PC, we can compute PCA projections of a
HapMap ENCODE region in a matter of seconds.  
However, there are non-trivial numerical issues with these
computations which makes them more difficult in our case.
What we are seeking is the correct combinatorial structure of 
the genotopes and their projections.  However, 
arbitrarily small round-off errors can change the
combinatorics of a polytope. An example is the 
standard $3$-dimensional cube:  if we
slightly perturb one vertex, then at least one square 
face will be broken into two triangles. 

To solve this round-off problem, 
we computed each PCA projection of our 
ENCODE genotopes using very high precision arithmetic.  
Whenever we observed a facet of a 
projected genotope nearly parallel to one of its neighboring facets, 
we merged the two facets together.  
This process gives the correct facets.

\begin{table}
\begin{tabular}{c|c|l|l}
$\ell$ & $\sigma_{\ell}$ (r131, m014) & ENr131 \, $f$-vector & ENm014 \, $f$-vector \\
\hline
$1$ & $696,\,611$ & $(\, 2\, )$ & $(\, 2\, )$ \\ 
$2$ & $117,\,76$ & $(\, 11,\, 11\, )$ & (\,10,\, 10\,) \\
$3$ & $91,\,64$ & $(\, 46,\, 132,\, 88\, )$ & $(41,\, 117,\, 78\, )$ \\
$4$ & $69,\,52$ & $(\, 82,\, 505,\, 846,\, 423\, )$ & $(90,\, 549,\, 918,\, 459\, )$ \\
$5$ & $67,\,41$ & $(\, 135,\, 1567,\, 4938,$ & $(136,\, 1582,\, 4988,$ \\
& & $\,\,\,\,\,\,\,\,\,\,\,\,\,\,\,\,\,\,\,\,\,\,\,\,\,\,\, 5840,\, 2336\, )$ &
$\,\,\,\,\,\,\,\,\,\,\,\,\,\,\,\,\,\,\,\,\,\,\,\,\,\,\, 5900,\, 2360\, )$ \\
$6$ & $57,\,37$ & $(\, 179,\, 3570, \, 18699, \,$ & $(\, 182, \, 3544, \, 18370,$ \, 
\\
& & $ \,\,\,\,\,\,\,\,\,\,\, 39142,\, 35751,\, 11917\, )$ & $\,\,\,\,\,\,\,\,\,\,\, 
38300, \, 34938, \, 11646\,)$\\
\end{tabular}
\vskip .2cm
\label{PCAtable}
\caption{Singular values and $f$-vectors of the $\ell$-th PCA projection of the 
ENCODE genotopes for regions ENr131 and ENm014, for $\ell = 2,3,4,5,6$.}
\end{table}

We computed the $\ell$-th PCA projection of the
two ENCODE genotopes for $\ell$ up to $6$.  For each projection we give
the $f$-vector, which records the number of faces of each dimension.
The results are shown in Table~1.
We note that the first few principal components explain most of the 
variation in the data, and that the $f$-vectors of the 
$\ell$-th PCA projections are quite similar between the two regions.
Figure 2 shows the 3-dimensional  PCA projection of the ENr131 genotope.
The $f$-vector is $(46,132,88)$ which means that the polytope has $46$
vertices, $132$ edges, and $88$ facets. Each vertex of the polytope is a
projection of one of the $269$ genotypes in the four 
populations, which are indicated by colors.

We found that the projected genotopes
give an excellent representation of the
four different populations, in the sense
that they correspond to four distinct
regions on the polytope boundary.
The only exception in Figure 2 is one of the
Japanese genotypes that occurs among the
Utah genotypes. While this may be a coincidence, it is noteworthy
that only the identification of Japanese in the genotyping process 
was solely based on self-reporting.

\section{Projections Onto Few SNPs}

 \begin{figure}
\begin{center}
 \includegraphics[scale=0.59]{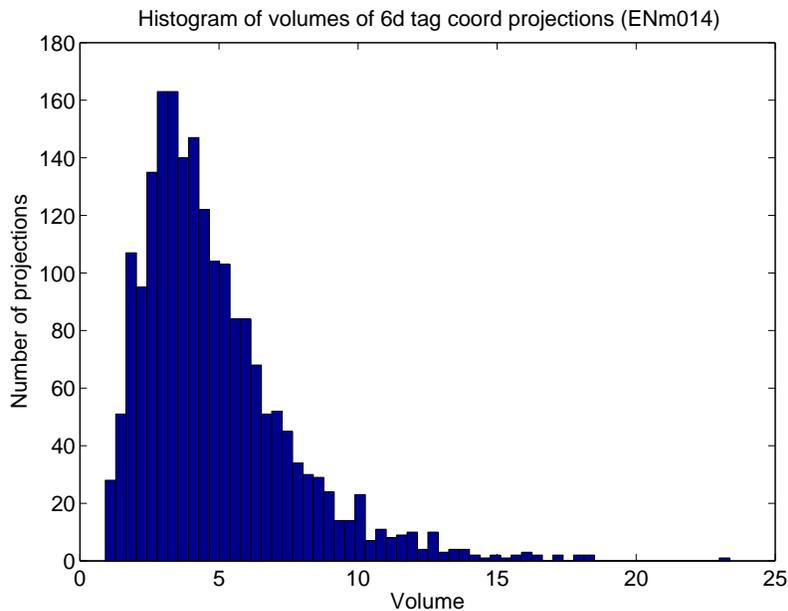}
\caption{Volumes of projections of the ENm014 genotope onto tag SNPs.}
\end{center}
\end{figure}

In this section we discuss coordinate projections of the
ENCODE genotopes. Such projections are relevant because of the
commonplace practice of selecting {\em tag SNPs} for genetics studies.
Such SNPs are subsets of the available SNPs that are, as much as
possible, in pairwise linkage disequilibrium. Thus, despite the fact
that $\ell$ tag SNPs capture less of the variation in the data than the
$\ell$th PCA projection, they
are useful in re-sequencing applications where it is desirable to sample
as few SNPs as possible.

The polyhedral analog of restricting analysis to tag SNPs is the
projection of the genotope onto the tag SNP coordinates.
Each individual projection onto $\ell$ SNPs is easy 
to compute, provided $\ell$ is not too big. What makes
the computation of all coordinate projections
 challenging is the combinatorial explosion
in the number of  $\ell$-element subsets of the $n$ SNPs. 
We did not attempt to exhaustively compute all 
of these projections. Instead we 
computed the projections of the two ENCODE genotopes onto 
tag SNPs selected for further study.

Using the software {\tt Hclust.R}
\cite{Rinaldo2005}, we chose 35 tags out of the 790 SNPs for region
ENm014,
and
109 tags out of the 1154 SNPs for region ENr131.  These rather small
sets
of tag SNPs still capture most of the variation in the data:    
for region ENm014, the sum of estimated variances of the original 790 columns 
is 108, and after projecting all columns onto the span of the 35 tag columns
(and an added column of ones), the sum of the 790 estimated variances is 95.  
Similarly for region ENr131, the original sum of estimated variances is 288, 
and becomes 274 after projecting.  Many of the SNPs not chosen
as tags were monomorphic in the HapMap data, 
or had low observed minor allele
frequencies. For example, in region ENm014 
there were 206 monomorphic sites and 331
with low minor allele frequency. {\tt Hclust.R} reported that 253 candidate
SNPs were considered
for region ENm014, and 726 candidate SNPs were considered for region
ENr131.  We then investigated random
samples of coordinate projections of the ENCODE projections onto $\ell$
tag SNPs for $\ell = 2,3,4,5,6$.  This computation of tens of thousands
of polytopes was accomplished by automating polyhedral software.
The packages we used are {\tt polymake} \cite{polymake} and {\tt iB4e} \cite{iB4e}.

Our geometric analysis suggests a new test for identifying informative SNPs.
For each coordinate projection, we compute the volume of the resulting projected
genotope. Since points in a genotope correspond to allele frequency vectors which can
be realized by populations over the genotypes, there is a natural probabilistic
interpretation of such volumes of genotopes. The {\em volume criterion}
seeks to identify the subset of $\ell$ SNPs which maximizes this 
polytope volume.

As an example of our coordinate projections data, 
Figure 3 shows the empirical distribution of the volumes of the 
$6d$ tag SNP projections of the ENm014 genotope. 
In our random sample of 2000 6d projections onto tag SNPs for region ENm014, 
the largest volume we observed was 23.36.  The projection attaining this
maximal volume is a genotope with 84 vertices and 377 facets, which is high
compared to randomly chosen 6d projections onto tag SNPs.  Moreover, this
particular 6d projection explains almost half of the variation in our data
matrix for region ENm014.  Out of the random sample of 2000 6d tag SNP 
projections, only 13 produced a higher sum of variances of the projected
columns.  We take this as strong empirical support for our proposed volume
criterion.

\section{Archetypal Analysis}

Archetypal analysis was introduced by Cutler and Breiman
\cite{Cutler1994} as an alternative to PCA. Its aim is to find low-dimensional
projections of the data points onto meaningful mixtures of the
high-dimensional points. Our data points in this section are
the rows of our data matrix $X$, i.e., the  
$k$ genotypes $x_1,\ldots,x_k \in \{0,1,2\}^n$. They
represent the individuals in the four populations.

Archetypal analysis finds {\em archetypes} that have the property
that when each genotype is replaced by its
nearest mixture of archetypes, the total least squares error is minimal.
More precisely, if $\ell$ is the number of archetypes to be found
(specified by the user), then the goal is to find archetypes
$z_1,\ldots,z_{\ell}$ together with $\alpha_{im}$ and $\beta_{mi}$ ($0 \leq
\alpha_{im},\beta_{mi}$ and $\sum_{m}\alpha_{im} = \sum_{i}\beta_{mi}=1$) such that
\[ z_m \quad = \quad \sum_{j=1}^k \beta_{mj}x_j, \qquad m=1,\ldots,\ell, \]
and 
\[ \sum_{i=1}^k||x_i-\sum_{m=1}^{\ell}\alpha_{im}z_m||^2\]
is minimized.

The benefit of archetypal analysis is that the archetypes have a useful and
meaningful interpretation. For the data studied here, the
archetypes are mixtures of genotypes. Thus the inferred archetypes can be interpreted as
representative populations for the measured genotypes.
 No efficient algorithm is known that guarantees
finding the $\ell$ optimal archetypes, but the alternating optimization
procedure suggested in \cite[\S 4.1]{Cutler1994} appears to perform well in practice. 

In our view,
computing archetypes from human variation data
may be useful for designing population based genetic
studies. In particular, the allele frequencies of the archetype
populations suggest sampling strategies for controls in
case-control studies, where it may be useful to sample a small
number of groups of controls whose allele frequencies match
the archetype populations.

We now explain our implementation of archetypal analysis.
The function to be minimized is
a large biquadratic polynomial, that is, a polynomial which is
separately quadratic in two groups of unknowns.
This biquadratic polynomial is the {\em residual sum of squares} (RSS)
whose derivation is given  in \cite[\S 4]{Cutler1994}.
Our problem is to minimize the residual sum of squares subject
to non-negativity constraints. This optimization problem can have many local
minima, and, in general, we cannot hope to find the
global minima. The heuristic of alternating optimization
for computing local minima works as follows.
If we keep one of the two sets of variables fixed, then
the objective function is just quadratic in the other set of
unknowns and we can easily find the global minimum.
We then fix those values and we allow the
other set of unknowns to vary, solving again a
quadratic optimization problem. Iterating this
procedure leads to a local optimum \cite[\S 5]{Cutler1994}.
This process can be repeated with many
different starting values to reach a local 
optimum that is eventually satisfactory.

 \begin{figure}
\begin{center}
 \includegraphics[scale=0.55]{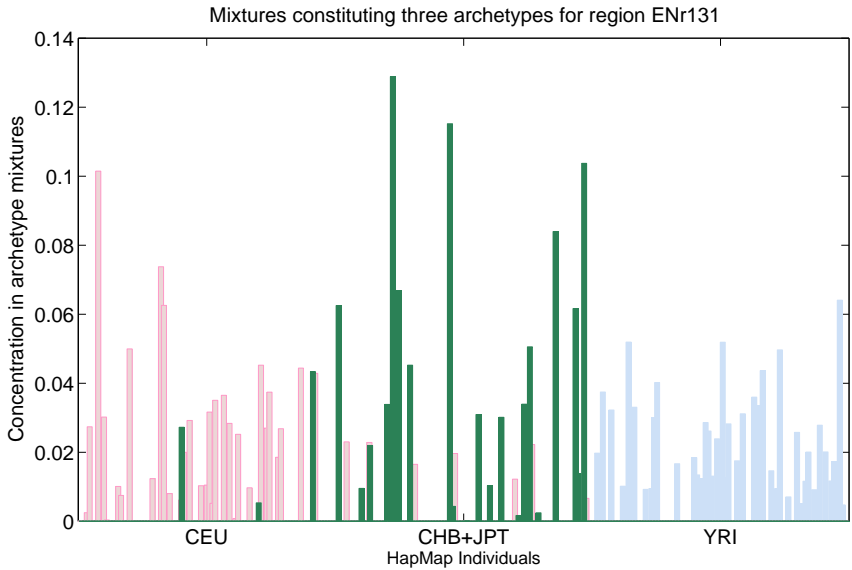}
\caption{Thearchetypes for region ENr131 showing the population structure.}
\end{center}
\end{figure}

We implemented this alternating optimization algorithm in MATLAB,
using the high-performance optimization package {\tt SeDuMi} \cite{Sedumi}
to solve the arising quadratic optimization subproblems.
As a heuristic to speed up computations,
we first restricted our attention to tag SNPs and computed archetypes for
these much smaller data sets.  We then used the obtained
archetypes as an initial guess for the archetypes in the
full-dimensional data.

We computed sets of three archetypes for ENr131 and ENm014.
The number $\ell = 3$ is of particular interest in our study since we advocate that 
archetypes should be interpreted as representative populations, 
and the HapMap data is 
derived from three populations (CEU, JPT+CHB, and YRI).  

Figure 4 depicts the three archetypes computed for region ENr131.  
By definition, each archetype can be expressed 
as a mixture $\sum \beta_i x_i$ where
the $x_i$ are the genotypes of the 269 HapMap individuals, 
and all $\beta_i \geq 0$, with $\sum \beta_i = 1$.  For each archetype, we 
plot the 269 coefficients $\beta_i$ as 269 bars in the figure. 
We plotted all three archetypes on the same figure, using a different
bar color for each archetype.  
We confirm that the three archetypes correspond to 
the three main geographic populations CEU, JPT+CHB, and YRI.  Although there 
is some mixing between CEU and JPT+CHB in two of the archetypes,
the third archetype is purely YRI and it is the only archetype
containing YRI. This is consistent with the fact
that YRI is an
outgroup among the three populations.

\section{Discussion}
As the amount of human variation data continues to increase, it is
becoming imperative to find representations of the data that are
informative and convenient for analysis. The human genotope offers one
such representation: a geometric description of the data that is useful for studying
population structure and epistasis. In this paper, we have computed
three low-dimensional projections of a subpolytope of the human
genotope. The projections are all compatible with the geometric
structure of the genotope, and are useful in different ways. 

In terms of
population structure, we believe that the results using archetypal analysis
are particularly interesting, and we were surprised at the natural
separation of the populations that emerged in the archetypes (Figure 4). 
In our computations we picked the number of
archetypes to be $\ell = 3$ based on information we had about the data. In
general, it is an interesting problem to determine, in a statistically
meaningful way, the ``optimal'' number of archetypes to use for an
analysis. We believe that information theoretic measures, such as the
Jensen-Shannon divergence, may be useful for this problem \cite{Grosse2002}.
The principal component projections in Section 4 demonstrate that
the genotope is a representation that is compatible with existing
approaches to dimensionality reduction. In particular, we have shown
that principal component projections can be applied to polytopes, and
not just points. Our results in Figure 2 offer a geometric
analog of Cavalli-Sforza's gene-based population analyses \cite{Cavalli-Sforza}.

We conclude by reiterating that our results are an initial first step
towards the construction of the human genotope. Next steps include the extension to
 more SNPs, association of phenotype data with the
genotypes so that epistasis can be studied, and an analysis of how the
genotope changes over time.

\section*{Acknowledgment}

This work was supported by in part by the Defense Advanced Research Projects Agency (DARPA) under grant HR0011-05-1-0057.


\end{document}